\documentclass[aps,prc,showpacs,twocolumn]{revtex4}
\usepackage{graphicx}
\usepackage{dcolumn} 
\usepackage{bm}
\usepackage{hyperref}
\usepackage{natbib}
\usepackage{amsmath}
\usepackage{multirow}
\usepackage{textcomp}
\begin{document}
\title{Allowed $\beta$-decay spectrum with numerical electron wave functions}
\author{Dong-Liang Fang\footnote{Current address: Institute of Modern Physics, Chinese Academy of Sciences, Lanzhou, 730000, China}}
\address{College of Physics, Jilin University, Changchun, 130012, China}
\address{Institute of Modern Physics, Chinese Academy of sciences, Lanzhou, 730000, China}
\begin{abstract}
Using numerical electron wave functions and state-of-the-art nuclear many-body methods, I evaluate the $\beta$-decay spectrum for typical decay channels of spherical nuclei. I check errors brought by various approximations used in deriving the analytical shape factors (the so-called Fermi Function) of allowed decay. I estimate the errors brought by the different electric charge distributions and give a way of estimation of $\beta$-spectra with available decay data of specific nuclei. I found that the traditional ways of approximating the electron wave functions by Fermi Function could be a very severe source of error for spectra simulation.
\end{abstract}
\maketitle
\section{introduction}
Reactor anti-neutrino spectra from nuclear $\beta$-decay fission products is important for measurement of various properties of neutrinos, and for the potential discovery of sterile neutrino. In Ref.\cite{Hub11}, with the inversion spectra technique, one constructs the reactor anti-neutrino spectra. Another method of constructing the neutrino spectra, the so-call {\it ab initio} method has also been developed\cite{Mue11}, where one sums the spectra of single decay branches together and get the final reactor neutrino spectra. One then finds that these constructed spectra deviates from measurements, this is the so-called reactor anti-neutrino anomaly\cite{MFL11}. One suggests that this could be the effect from the existence of extra sterile neutrinos which don't participate weak interactions. To draw such conclusion, one must make sure the prediction of neutrino spectra is accurate enough. While the {\it ab. initio} approach is somehow restricted by the incompleteness of current nuclear database \cite{SJM15}, both {\it ab initio} and reversion methods rely heavily on accurate simulations of $\beta$-decay neutrino spectra for specific decay channels. This somehow needs to be carefully checked with modern nuclear structure calculations. As a first step, I consider leading order contributions (allowed $\beta$-decay) to decay width and $\beta$- as well as neutrino spectra.

It is known in standard model, charged weak interaction is mediated by the SU(2)$_L$ charged gauge bosons $W^{\pm}$ which gain mass through Higgs mechanism. Due to the heavy mass of W gauge bosons, the weak interaction at low energy scale behaves like four-Fermion contact interaction(The simplest form is the so-called Fermi theory\cite{Fer34}):
\begin{eqnarray}
\mathcal{L}_{W}&=& \frac{G_F \cos \theta_{C}}{\sqrt{2}}\bar{\psi}_{p}(\vec{x}) (g_V \gamma_\mu -g_A \gamma_5\gamma_\mu) \psi_{n}(\vec{x}) \nonumber \\
&\times& \bar{\psi}_{e}(\vec{x}) (1-\gamma_5) \gamma_\mu \psi_\nu(\vec{x}) 
\end{eqnarray}

With certain approximations, these currents were then reduced to non-relativistic form to suit the nuclear structure calculations, this was done decades ago by various authors\cite{Wei61,BBC77}. With further approximations (will be mentioned below) made subsequently, one then got analytical expressions for decays width of nuclei\cite{Wei61} with point electric charge distribution. Later on, one simulated the effect of finite nuclear size with expansion method where one expands the electron wave functions on quantities such as $\alpha Z$, $k R$ and $m_e R$\cite{Bue63}(for a recent review see Ref.\cite{HSB17}). These analytical expressions give errors about several percents to the exact solutions, this is good for half-life calculations, but when going to spectra simulation, these errors must be considered seriously.  So in this work, instead of analytical derivations of $\beta$-spectra, I use numerical electron wave functions, and combine with modern nuclear many-body approaches, this could us give quantified estimation of errors of these analytical spectra. This will also help reduce the errors of neutrino spectra and helps to clarify the probability of existence of sterile neutrino in reactor.

This work is arranged as follows: I present the formalism at section II, and the choice of parameters in section III, later on results of static potential of point charge and distributed charge in section IV with also the estimation of errors presented.

\section{Formalism}
The decay width of $\beta^-$ decay for spherical nuclei can be expressed in an exact form\cite{Wei61}:
%\begin{widetext}
\begin{eqnarray}
%\lambda&=&\frac{\ln 2}{t_{1/2}} \nonumber \\
\lambda&=&\frac{G_{\beta}^2}{(2 J_i+1)2\pi^3}  \sum_{K} \sum_{\kappa_e \kappa_\nu} \frac{1}{2}\int \{ \sum_{Ls} \int  \sqrt{\frac{ 4\pi}{2K+1} } \nonumber \\
&\times& \langle\langle J^{\pi'}_f ||(1+\frac{g_A}{g_V}\gamma_5)T_{KLs} || J_i^{\pi} \rangle\rangle  \\
&\times& \sqrt{4\pi}\langle \langle \phi_{\kappa_e}(Z) ||(1+\gamma_5)T_{KLs}|| \phi_{\kappa_\nu}\rangle \rangle r^2 dr \}^2 p^2 q^2 dp \nonumber
\end{eqnarray}
%\end{widetext}
Here and after the double bras and kets integrate over the angular parts only for leptons and nuclear wave functions. And $T_{KLs}$ is the decomposed component of weak current over the angular momenta derived by \cite{Wei61} and can be written in a compact form:
\begin{eqnarray}
T_{KLs}^\mu=C_{L M s \mu-M}^{K\mu} Y_{L M} (\gamma_5 {\bf \sigma}_{\mu-M})^s
\end{eqnarray}

For the allowed case, we have the leading contribution $K=1$, $L=0$ and $s=1$ (Other operators may contribute, but a naive analysis suggests that their contributions are negligible), this reduces to the non-relativistic allowed beta decay operator as:
\begin{eqnarray}
\mathcal{O}_{GT}=\gamma_5 T_{101}=\frac{1}{\sqrt{4\pi}} \sigma
\end{eqnarray}
at the leading order, with the induced current ignored. 

Then the lepton part has the form\cite{BBC77}:
\begin{eqnarray}
F_{\kappa_e,\kappa_\nu} (p,q,r)&\equiv&\sqrt{4\pi} \langle \langle \phi_{\kappa_e}(Z) ||(1+\gamma_5)T_{101}|| \phi_{\kappa_\nu}\rangle\rangle \nonumber \\
&=&g_{\kappa_e}(Z)[j_{l_\nu}G_{101}(\kappa_e,\kappa_\nu)]+\cdots
\end{eqnarray}
Here $g_{\kappa_e}$ and $f_{\kappa_e}$ are the upper and lower components of electron wave functions with quantum number $\kappa_e$ defined in\cite{Wei61}. They are solutions of Dirac equations for a specific central potential. And from above expressions, one finds the following relation $F_{\kappa_e,\kappa_\nu}=F_{\kappa_e,-\kappa_\nu}$, this could simply the calculations with less terms.

Meanwhile according to the selection rule that parity change of allowed decay is +1, this then implies only the axial vector current contributes to the nuclear parts, altogether for allowed decay the decay width has the form:
\begin{eqnarray}
\lambda_{GT}&=&\frac{G_\beta^2 }{( 2 J_i+1)2\pi^3} \sum_{\kappa_e, \kappa_\nu>0} \int_{m_e}^Q d\epsilon\quad p \epsilon (Q-\epsilon)^2  \nonumber \\
&\times& \{ \int  \frac{1}{\sqrt{3}} \frac{g_A}{g_V}  \langle\langle J^{\pi'}_f || \sigma || J_i^{\pi} \rangle\rangle F_{\kappa_e,\kappa_\nu}(p,Q-\epsilon,r) r^2 dr \}^2  \nonumber \\
\label{dwgt}
\end{eqnarray} 
Here $Q$ is the mass difference of states of parent and daughter nuclei, and $\epsilon$ is the electron energy. With numerical electron wave functions and modern many-body approaches, we could get accurate decay width as well as the electron spectra from above formula. So I define the following nuclear transition matrix elements:
\begin{eqnarray}
B_{\kappa_e,\kappa_\nu}(p,q)= \langle J^{\pi'}_f| F_{\kappa_e,\kappa_\nu}(p,q,r) \sigma |J^{\pi}_i \rangle 
\end{eqnarray}

Then I can get the differential decay width from eq.(\ref{dwgt}) to get the $\beta$-spectra, the neutrino spectra can be derived in a similar way, I will not present the results of latter in this work. And the transition strength $B_{\kappa_e,\kappa_\nu}$ can be obtained from nuclear many-body calculations:
\begin{eqnarray}
&&B_{\kappa_e,\kappa_\nu}(p,q) \nonumber \\
&=& \sum_{pn}\langle J^{\pi'}_f|| [c_p^\dagger c_{\tilde{n}}]_{1^+} ||J^{\pi}_i \rangle \langle p|| F_{\kappa_e,\kappa_\nu}(p,q,r) \sigma ||n \rangle 
\end{eqnarray}
Here $\langle J^{\pi'}_f|| [c_p^\dagger c_{\tilde{n}}]_{1^+} ||J^{\pi}_i \rangle$ is the one-body reduce matrix element. The detailed derivation of these terms for QRPA and Shell model calculations used in this work can be found in {\it e.g.} \cite{FBS13}. This leads to the final expression of the differential decay width of this work:
\begin{eqnarray}
\frac{d\lambda_{GT}}{d\epsilon}=\frac{m_e^5}{\ln 2 C} \sum_{\kappa_e,\kappa_\nu} B_{\kappa_e,\kappa_\nu}(p,Q-\epsilon) p \epsilon (Q-\epsilon)^2 
\label{dw}
\end{eqnarray}
Here $C \approx 6170s$ is a constant widely used in literature for $\beta$-decay. And we can define $S(\epsilon)=(d\lambda/d\epsilon)/C$ as differential decay width in unit of $C^{-1}$ in this work.

In analytical calculations for allowed beta decay\cite{Wei61}, several approximation is used. At first, one used the so-called neutrino long wave approximation($\nu$LA), where only s-wave neutrino is taken into account with its value being unity:
\begin{eqnarray}
j_0(qr)=1\quad j_l(qr)=0,\quad l\ne 0 
\end{eqnarray}
The same is used for electron, but because of the coulomb potential of nuclei, one adopts following approximation: 
\begin{eqnarray}
g_{\kappa_e}(kr)=g_{\kappa_e}(kR)\quad f_{\kappa_e}(kr)=f_{\kappa_e}(kR)
\end{eqnarray}   
Here R is the mean nucleus radius with empirical values $R=1.2 A^{1/3} $fm or the so-called Elton formula $R=1.12 A^{1/3}+2.43A^{-1/3}-6.56A^{-1}$ fm. And $g_{\kappa}(kR)$ and $f_{\kappa}(kR)$ refer to wave functions of electrons with momentum $k$ at nuclear surface,  hereafter, this refers to surface approximation(SA). The same SA can be applied to neutrino if we consider neutrino wave functions beyond LA.

These approximations simplify the decay width to the form, {\it i. e.}  with electron SA and $\nu$LA:
\begin{eqnarray}
\lambda_{GT}&=&\frac{G_\beta^2 }{( 2 J_i+1)2\pi^3}  \int_{m_e}^Q d\epsilon \  p \epsilon (Q-\epsilon)^2 \nonumber \\
&\times&   \{ \int g_A \langle\langle J^{\pi'}_f || \sigma || J_i^{\pi} \rangle\rangle r^2 dr \}^2 (g_{-1}^2(pR)+f_1^2(pR))  \nonumber \\
& =&(\ln2 C)^{-1} f g_A^2 B(GT)
\end{eqnarray}
Here $B(GT)=|\langle f | \sigma |i \rangle|^2$ is the so-called Gamow-Teller strength. And $f$ is the so-called phase space factor which can be expressed as following form for analytic electron wave solution of point charge coulomb potential with only leading terms\cite{Fer34}:

%With derived analytical point charge solution of electrons, this expression further 
%With these approximations, after substitution of $G_{101}(\kappa_e,\kappa_\nu)$, we come to the final expression:
%\begin{eqnarray}
%\lambda_{GT}&=&\frac{G_\beta^2 }{( 2 J_i+1)2\pi^3}  \int_{m_e}^Q d\epsilon \  p \epsilon (Q-\epsilon)^2 (g_{-1}^2(pR)+f_1^2(pR)) \nonumber \\
%&\times&   \{ \int g_A \langle J^{\pi'}_f || \sigma || J_i^{\pi} \rangle r^2 dr \}^2 
%\end{eqnarray}
%We find that now the integration over momentum and coordinate space can be well separated and above expression can lead to a much common expressions:
%\begin{eqnarray}
%\lambda_{GT}=\ln 2/t^{1/2} = f g_A^2 B(GT)/8896 s^{-1}
%\end{eqnarray}
%Here $B(GT)$ is Gamow-Teller strength defined as $B(GT)=\frac{1}{2J_i+1}|\langle f ||\sigma || i \rangle |^2$ (here brackets are with the normal definition).
%And $f$ is the phase space factor:
%\begin{eqnarray}
%f= \int_{1}^W d\omega \  p \omega (W-\omega)^2 (g_{-1}^2(pR)+f_1^2(pR))
%\end{eqnarray}

\begin{eqnarray}
f= \int_{1}^W d\omega \  p \omega (W-\omega)^2 F_0(Z,W)
\end{eqnarray}
Here $W$ and $\omega$ are electron energies in unit of electron mass.

As we can see above, by deriving the $\beta$-decay rates and $\beta$ spectra, analytically, one neglected the electromagnetic interactions between nuclei and electrons. And the wave functions are usually treated on the base of a point charge coulomb distorted one, these will not largely affect the half-life largely, but when going to the $\beta$ or neutrino spectra, precision of a few percents is needed for simulations such as reactor neutrino flux, such effects could play important roles.
%, so with up-to-date many body approaches, we make numerical integrations in the coordinate space for nuclear wave functions as well as numerical electron wave function for specific nuclear coulomb potential for nucleus with detailed charge distributions to study the errors from analytic calculations.

\section{Parameters}
In this work, for choices of decay branches, I adopt different types of typical branches to understand the effects of different parameters such as Z, A and nuclear radius. Since in analytical calculations, the electron wave functions are expanded over $\alpha Z$ and $k R$ \cite{Bue63}, I take the lighter and heavier nuclei which can be well described with QRPA or Shell Model methods. And the upper limit of momentum $k$ is associated with decay Q-values, therefore I take decay branches with small and large Q-values. With these choices, one could make estimations of errors of analytic calculations. So the four chosen branches are: I) $^{50}$Ca with $Q=3.621$MeV, II) $^{58}$Ca with $Q=11.461$MeV, III) $^{120}$Cd with $Q=2.282$MeV and IV) $^{132}$Cd with $Q=8.650$MeV. All these nuclei are assumed to be of spherical shapes so that the spherical symmetry is enforced, especially for calculations of lepton parts. If sorted by the order of Q values, we find that decay branch III has the lowest Q values and II has the largest while branch I is slightly larger than III and IV in between I and II.

For nuclear part, different approaches are used for different isotopes due to their properties. For the closed shell Ca isotopes, since $Z=20$ is the magic number, nuclear shell model can be ideal for the simulation while for heavier nuclei such as Cd isotopes, we need to resort to some kinds of less exact methods such as QRPA. Thus in this work, the Ca isotope results are obtained with shell model calculations with the GXPF1a Hamiltonians\cite{HOB02} and NuShellx@MSU\cite{BR14}. While the Cd results are obtained from spherical QRPA calculations with realistic forces\cite{FBS13}. For QRPA calculations, I used the CD-Bonn potentials and the parametrization is as in \cite{FBS13}. For both calculations, since we are more interested in the spectra rather than reproducing the half-lives, so $g_A$ is simply taken as unity, the $q^2$ dependent quenching will slightly affect the spectra and will be discussed in future works. When a realistic values of $g_A$ have been adopted, one does find good agreements for half-life calculations\cite{HOB02,FBS13}. 

For the lepton part, I use for neutrino wave the partial wave decomposed from the plane wave solutions without account of any weak charge modifications. And for electrons, the partial waves are obtained from solving the Dirac equations with specific static electric potentials. Four potentials are used for this calculations, one is from a point like charge distributions and other three for finite nuclear size charge distributions. As it is known, the potential for a point particle is simply $V(r)=-Ze/r$. For the case with a finite nuclear size, we assume three different distributions as suggested in\cite{BB70} for sake of comparison: 
I) the uniform charge distributions where the electric charge is uniformly distributed inside a sphere with radius defined by the Elton formula\cite{Elt58};
II) the Fermi distributions as expressed in eq.(9) of \cite{BB70}. and with parameters therein;
II) the modified Gaussian distributions expressed as eq.(12) in \cite{BB70} with parameters therein. By comparing with the electron-nucleus scattering experiments\cite{DDD87}, I find the latter two parametrizations agree with experiments quite well.

The static potentials for these charge distributions are then obtained by solving the Coulomb equations. These Coulomb potential are used to get the final electron wave functions by solving Dirac equations. This is done by the numerical subroutines called {\it RADIAL}\cite{SFW95}. In my calculation, I set the numerical accuracy of the subroutine to be $10^{-10}$ considering both the accuracy and efficiency.

\section{Results}

\subsection{Point Charge distribution}
\begin{figure*}
\includegraphics[scale=0.6]{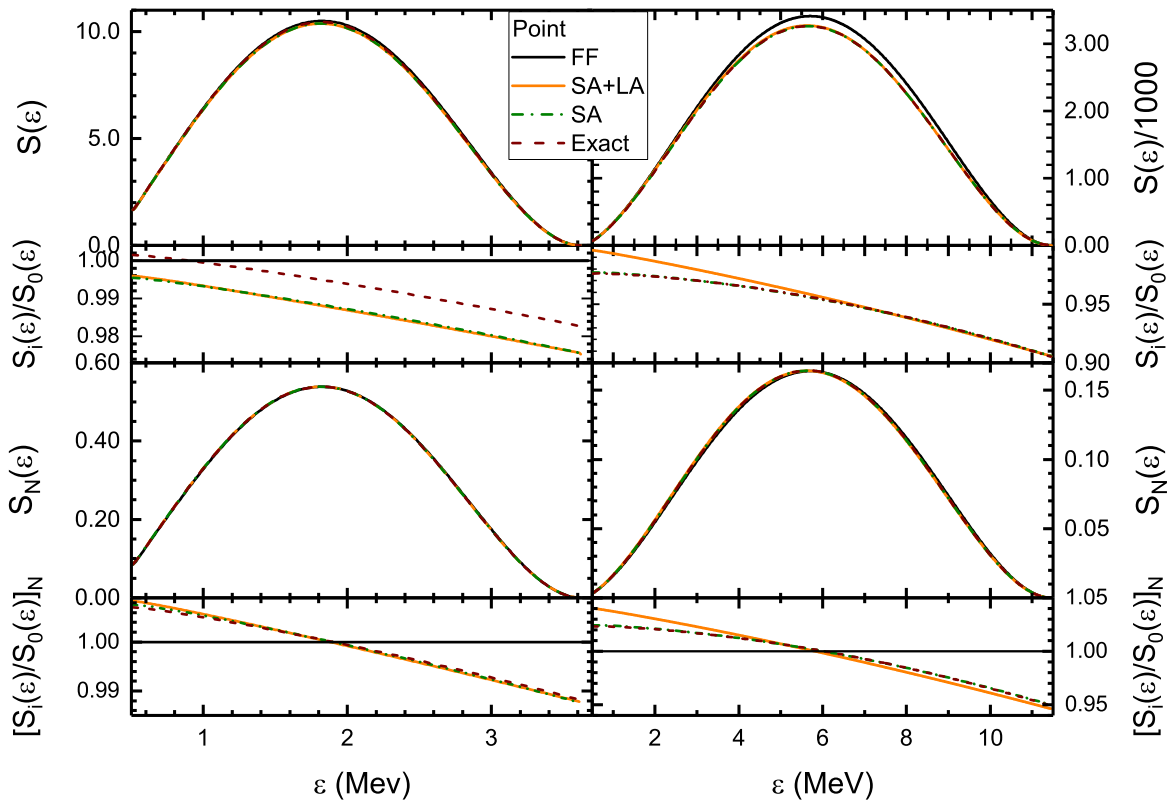}
\caption{(Color online) Spectra for Ca isotopes with the decay branches specified in text. Here, FF refers to traditional Fermi Function, point refers to usage of exact point charge solution instead of Fermi function, $\nu\quad WF$ refers to usage of true neutrino wave functions instead of long-wave approximations, $Exact$ refers to full nuclear structure calculations of $\beta$-decay width as explained in text. The subscript $N$ refers to normalized spectra.}
\label{caist}
\end{figure*}

\begin{figure*}
\includegraphics[scale=0.6]{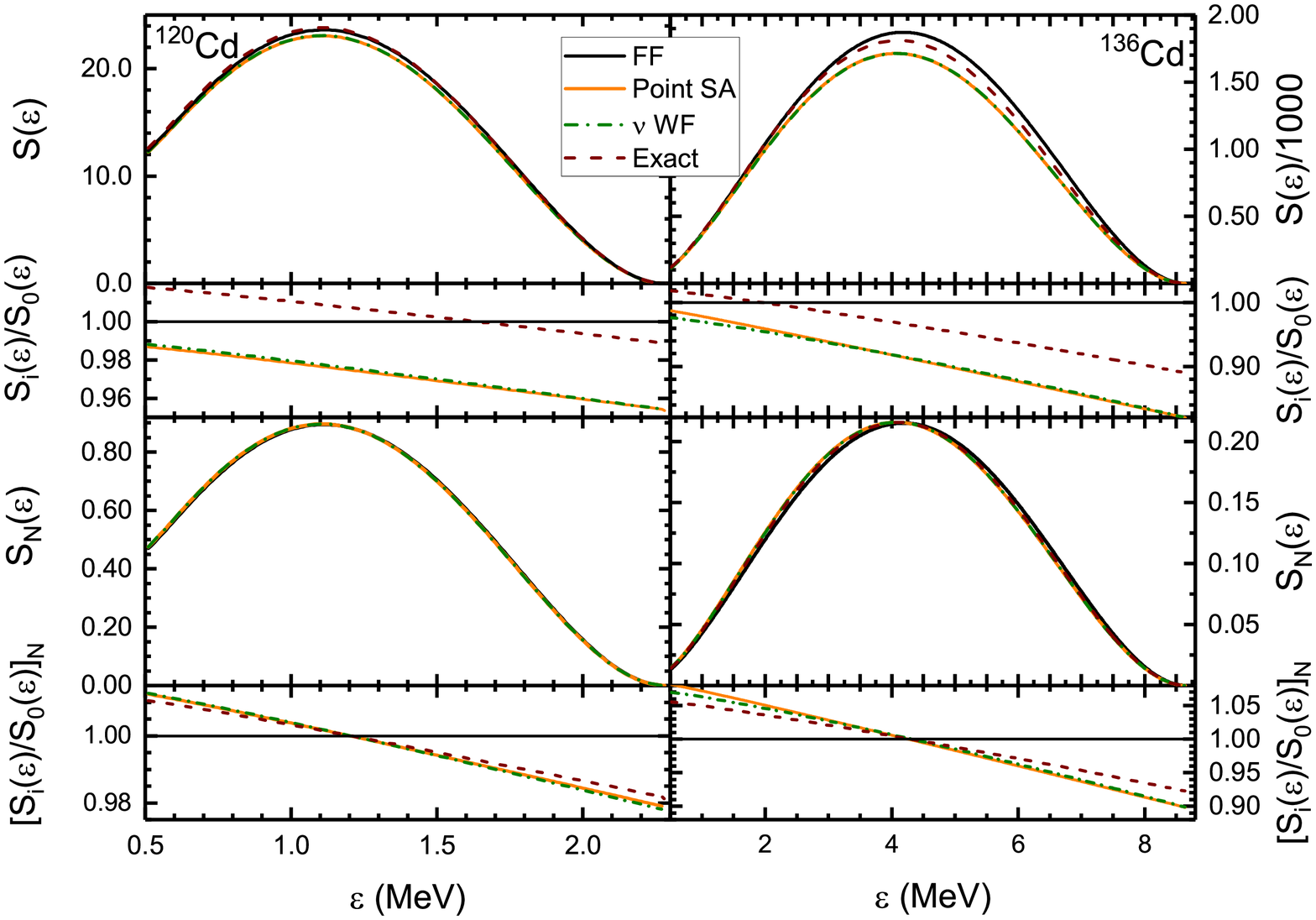}
\caption{(Color online) Similar as fig.\ref{caist} but for decay channels of chosen Cd isotopes.}
\label{cdist}
\end{figure*}

In this part, I first consider the case that the nuclear charge is concentrated on one point. In this occasion, one could have exact solutions\cite{You59}. This can be used as good calibration of the {\it RADIAL}\cite{SFW95} package for current calculation, I find that the numerical solution from {\it RADIAL} achieves very good precision even at very small radius.

\begin{table}[htp]
\begin{center}
\caption{The decay width (in unit of $C^{-1}$) for the four branches with different approximations. }
\begin{tabular}{|c|cccc|}
 \hline
 		&\multicolumn{3}{c}{Decay branches}& \\
		&     I 	& II 		& III		& IV		\\
\hline  
Fermi Function		& 559.588	 & 597605 & 757.463 & 249696  \\
Point SA+LA($\nu$)	& 552.720 & 572146 & 738.360 & 227923  \\
Point SA+$\nu$ WF 	& 552.811 & 569950 & 739,109 & 227836  \\
Exact treatment 	& 556.487 & 570050 & 762.992 & 240915 \\
\hline
\end{tabular}
\end{center}
\label{tab:point}
\end{table}

\subsubsection{Fermi Factor and point-like charge potential Solution}
The exact solution of Dirac equations with point-like coulomb potential(Hereafter PCWF) have been obtained analytically ({\it e.g.} \cite{SFW95}), and to facilitate the actual calculation, one uses the Fermi Factor for the derivation of the phase space factors as well as spectra\cite{Fer34}. Not so much literature concerns the accuracy of such simplifications especially for heavier nuclei which are produced in the reactor as fission fragments. So at first I check how the approximation of FF to exact solutions affects the results providing with the neutrino long-wave-length approximation(LA) and electron wave functions constant inside nucleus. The electron wave functions take the value at the nuclear surface, namely the surface approximation(SA). For the nuclear radius, I adopt the Elton formula\cite{Elt58} in this work. In Table \ref{tab:point}, as expected, one finds a deviation of the decay width from Fermi Function to numerical surface electron wave functions and this deviation increases with larger $Z$ or decay energies $Q$. For decay branch I, the deviation is less than 1\%, but for II the deviation increase to about 4\%. For a larger Z (decay channel III), even with smaller Q values to that of branch I, the deviation is about 2\%. And the deviation is approaching 10\% for large Z and large Q (decay branch IV). This result is as expected, as when one derives the Fermi Function, the assumption of small $\alpha Z$ and $kR$ is used and this may be no longer valid for fission fragments in reactors.

From Fig.\ref{caist} and \ref{cdist}, I find that the actual values of differential decay width for PCWF is smaller than that predicted by Fermi Function and the deviation between these two becomes larger as electron energy or momentum increases. The deviation depends heavily on Z and nuclear radius( nuclear radius in this case depends on the nuclear mass number A  and also partially related to Z, therefore indirectly on Z and A). For Ca isotopes, with electron at a typical energy about 2$MeV$, the deviation is about 1\%, this value increase to 3-4\% for Cd isotopes; at the electron energy of about 8$MeV$ the deviation increase from 5\% for $^{56}$Ca to 10\% for $^{134}Cd$. This also changes the shape of decay spectra, more strength has been shifted to lower energy, therefore one observes increased strength at low $\beta$ energy and less strength at high $\beta$ energy region. These deviations are at the order from 1\% to 10\% for the four decay branches at the head or the tail of the spectra in my calculations. Again, heavier the nuclei are, fiercer the changes are.

\subsubsection{beyond Neutrino long wavelength approximation}
In most $\beta$-decay calculations, one considers only contributions from s-wave neutrino with the long-wave-length approximations, in this work I include contributions also from p-wave and even higher angular momentum. My calculation shows that only s-wave and p-wave with total angular momentum $1/2$($s_{1/2}$ and $p_{1/2}$) components give decent contributions, even for heavier nuclei, the higher momenta partial waves give negligible contributions. In Table.\ref{tab:point}, I find that modification brought by the exact neutrino wave functions to the decay width is relatively tiny, smaller than 1\% for all branches. 

On the other hand, one also finds that the inclusion of numerical neutrino wave functions leads to increase of decay width for some nuclei but decrease for others. A closer look shows that for smaller Q, the decay width increases, regardless of the atomic numbers. To understand this, one should revisit the expressions in eq.\eqref{dwgt}.  Our treatment beyond the neutrino LA reduces the s-wave contributions (The s-wave components at the nuclear surface $j_0(kR)$ is always smaller than unity) and increases the p-wave contributions (which is zero under LA) to the decay width. The smaller the values of $k$, the closer neutrino wave functions s-component to 1. The s-wave reductions and p-wave enhancements compete with each other, if $k$ is small, the enhancements are dominant and with large enough $k$, reductions dominate. This could easily been observed from the electron spectra, at the end points of the spectra which correspond to the zero energy of neutrino, the two lines with and without $\nu$LA converge and then with decreasing electron energies, we see enhancements of decay width for curves without LA, and at certain points the deviations of these two curves begin to decrease, and if the $Q_\beta$ of the branch is large enough, the two curves meet again and then start the reduction from contributions beyond LA. For decay branch III, the exact curve is always above the LA curve as $Q_\beta$ is small, for other branches, we could find the two curves meet again. This explains the decay width difference with and without neutrino LA.

Such behavior leads to the distortions of normalized $\beta$-spectra instead of the shift of strength, but the distortion is dependent on Q. For decay branches with small Q values, distortions for high electron energy end is observed, while for large Q values, significant changes appear at low electron energy end. The magnitude of the corrections shows slight Z dependence. But this correction is overall small, only at the magnitude of $1\%$ at the spectra ends.

\subsubsection{Interference between the nuclei and leptons}
And for point particle potential, I here compare the full numerical nuclear structure calculations with the above surface approximations(SA). The SA can be a reasonable one based on the assumption, that lepton wave functions are nearly constant inside nucleus, this can be true if nuclear radius is small. On the other hand, this could also work if the emitted leptons are concentrated on the nuclear surface. On the nuclear structure side, this can be realized by the case such that the GT decays are dominated by transition from single particle orbitals to single particles orbitals both near the fermi surface and they are always with a mean radius close to nuclear radius.

If we take into consideration of the radial dependence of lepton wave functions explicitly, one will find modification of decay width, whose magnitude depending on details of the decay branches. In Table.\ref{tab:point}, I present the results from above exact treatment with modern nuclear many body approaches and those with SA separating the nuclear and lepton parts. For both cases, the exact neutrino wave functions (exact radial dependence of wave functions or with neutrino wave function SA) are taken into account. For most decay branches, such a treatment leads to a increase of decay width. To understand this, we should be aware that the major contributions of electron wave functions are from s-wave, it is decreasing functions of radius when the radius is much smaller than electron wave length. Therefore, current results suggests that, most of Gamow-Teller $\beta$-decay happens somewhere beneath the surface, but this is an averaging effect, since calculations\cite{LBF14} show that the low energy Gamow-Teller transitions could be collective ones with participating of many single particle orbitals inside the nucleus. 

For the differential decay width or the energy dependent decay intensity, I find  a systematic increase of differential decay width compared the spectra with SA, such behavior give us clear evidence that most likely spatial location of emitted leptons are somewhere inside the nuclei near the surface. And the average spatial location of emitted electrons is somehow momentum dependent, this can easily be deduced from the slopes of the curves of the ratio of these decay rates to that of the Fermi Functions. If the average spatial location is momentum independent, then the differential decay width should be proportional to each other for the case of SA and exact treatment, but the green and brown curves in panel (b)'s of fig.\ref{fns} show the opposite especially for large Z and Q. Unlike electron wave function SA, the enhancement of differential decay width for exact treatment may sometimes (Branch III) be even larger the Fermi function shape factors.

The pattern of normalized spectra are therefore being distorted, exact calculations shift more strength to the high energy side. The exact spectra are therefore lying between the Fermi Function and electron WF SA curves. This may suggest we may constraint the actual spectra by these two shape factors for point charge approximations. And this will be discussed for the finite size nuclear charge distributions in following sections.

And at last, let's look at the exception of the calculation, that is of decay branch II. In this case, one finds barely no deviations between the surface approximation and exact treatment (less than 1\textperthousand for decay width). I check the shell model calculations find this decay branch is dominated by the transition $f_{5/2}\rightarrow f_{7/2}$, these two orbitals lies near the fermi surface of neutrons and protons or equivalently at the nuclear surface, so for this decay branch, the emitted electrons and neutrinos are concentrated at the nuclear surface, making the surface approximation works perfectly. This again implies the importance of nuclear structure information on precise $\beta$ spectra calculations.

\subsection{Finite size of nuclear charge}

\begin{figure*}
\includegraphics[scale=0.65]{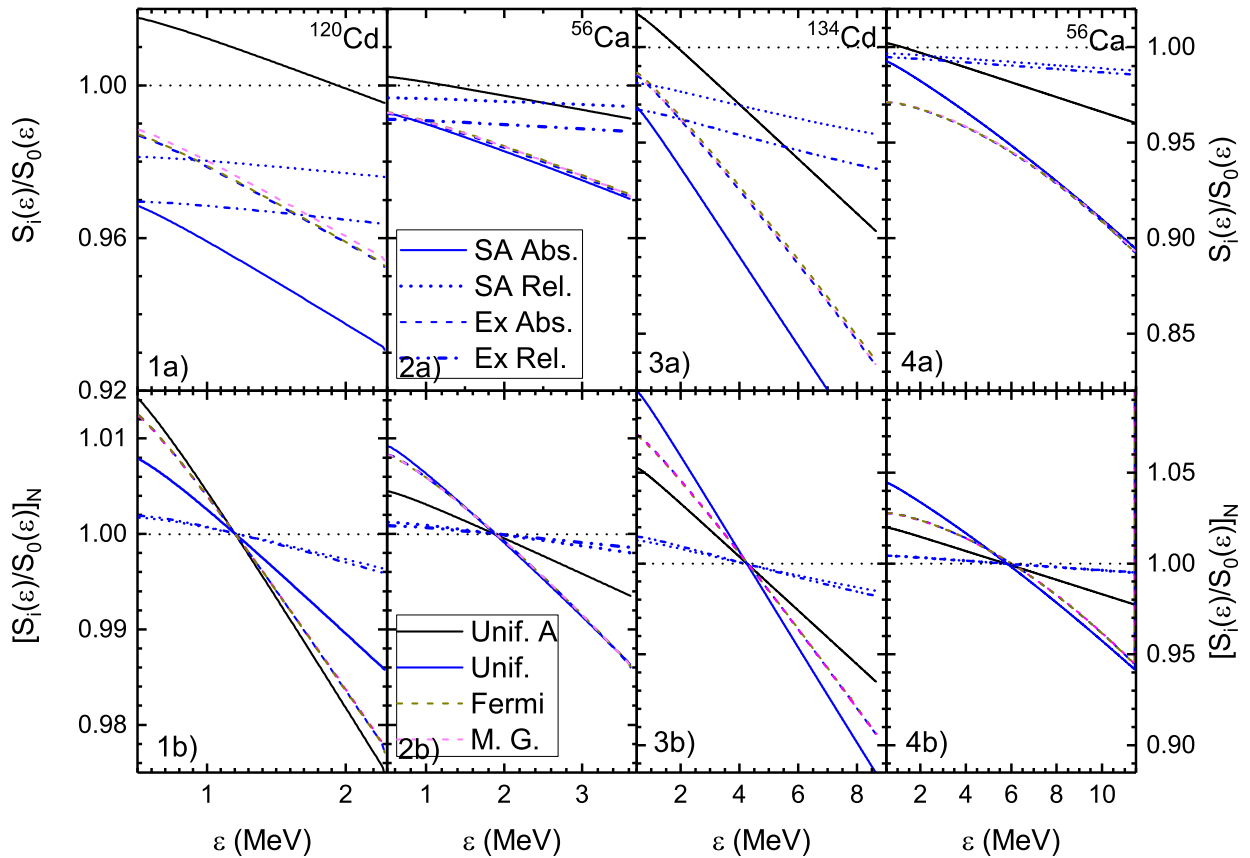}
\caption{(Color online) The differential decay width and normalized spectra for the four decay branches (indicated in text). Here Unif. means calculations use uniform charge distribution (With A refers to analytical), "M.G." and "Fermi" mean Modified Gaussian and Fermi distributions respectively as explained in the text. "Ex" means the exact calculations of decay width. Here "abs." and "rel." means absolute and relative corrections explained in the text.}
\label{fns}
\end{figure*}

In this part, I perform the $\beta$-decay calculations with the solutions of electron Dirac equations with different static electric potentials from different charge distributions. The change of the electrostatic potential from point charge to distributed charge shifts the wave functions inwards, this means generally reductions of decay widths as I will show.

\begin{table}[htp]
\begin{center}
\caption{The decay width (in unit of $C$) for the four branches with different nuclear charge distributions and treatment of finite nuclear size. Here SA refers to Surface approximation for electron wave functions and LA($\nu$) refers to the long-wave length approximations for neutrinos. }
\begin{tabular}{|c|cccc|}
 \hline
 		&\multicolumn{3}{c}{Decay branches}& \\
		&     I 	& II 		& III		& IV		\\
\hline  
Uniform analy.			& 558.331 & 587235 & 764.819 & 241316  \\
Uniform SA+LA($\nu$)	& 550.433 & 567881 & 723.287 & 220849  \\
Uniform Exact			& 550.829 & 564669 & 738.352 & 229653  \\
\hline
Fermi SA+LA($\nu$)		& 550.515 & 568056 & 723.970 & 221163  \\
Fermi Exact			& 551.088 & 564959 & 739.546 & 230193  \\
\hline
MG SA+LA($\nu$)		& 550.409 & 567850 & 723.172 & 220821  \\
MG Exact				& 550.992 & 564735 & 738.522 & 229720  \\
\hline
\end{tabular}
\end{center}
\label{fns}
\end{table}
%\subsubsection{Charge distributions and $\beta$-spectra}
The simplest charge distribution model is the uniform distribution and its corresponding electrostatic potentials can be expressed analytically\cite{BB70}. However, analytical electron wave function for this case is not available. Therefore, people may resort to different treatment: wave functions as Taylor expansion of several variables\cite{BB70,Wil90} or numerical ones\cite{SMN16}, the corresponding decay widths with these treatment are compared in this work. 

Analytical corrections of uniformly distributed nuclear electric charge is presented in\cite{Wil90}, where a parametrization based on $\alpha Z$ expansion of electron wave functions following\cite{Bue63} with a global fittings is used.  So here I first compare their results with numerical results with finite charge size wave functions(FSWF) under SA+$\nu$LA. What I found is that, deviations from the analytical and numerical calculations are observed. The analytical results predict larger decay width, and the deviations increase as electron energies increase. Generally, the analytical results predict the right trend of the change pattern of decay width relative to the point charge case if we ignore their magnitude. The absolute reduction of decay width of  electron WFs with distributed charges relative to FF varies from several percents to nearly  ten percents, depending on nucleus and Q values. The partial decay width reduces especially drastically for high electron momenta which can reach as high as 20\% for the example of decay branch IV with electron energy larger than 7MeV.

The formula in \cite{Wil90} is actually relative correction to PCWF which is approximated by the Fermi Function in previous publications. So to make reasonable comparison between the numerical and analytical corrections, one should look into the relative modifications from the numerical shape factors under SA+$\nu$LA to the numerical shape factor of PCWF with the same approximations, as well as those without any approximation under the two potentials. These are shown as the dotted and dashed-dotted curves in fig.\ref{fns}. I find that the size of finite nuclear size correction is usually smaller than that of analytical estimations. This is probably due to the fact that the parameters of analytical corrections are fitted globally\cite{Wil90} and these deviations may appear to be nucleus dependent, on the other hand, the Taylor expansion may become less accurate if $\alpha Z$ and $k R$ grow larger. However, the corrections show the same trend as the numerical ones that with increasing electron energies, the differential widths got reduced relative to the PCWF cases. This leads to shifts of decay intensity to the low electron energy end. So our new results with uniformly distributed nuclear charge agree with previous analysis\cite{Hub11}, the absolute magnitude differs from old results. In fact, the new results show that the relative correction from finite nuclear size is smaller than previous expectation\cite{Hub11} relative to exact point charge shape factor. But corrections relative to Fermi Functions can be well compared to other results for final spectra. The relative corrections due to finite electric charge size are smaller than 2\% for all the branches while the absolute corrections to Fermi Function can be as large as 10\% for specific branches, therefore, usage of exact PCWF instead of Fermi Function in the calculation could effectively improve the results without FSWF. 

The inclusion of exact neutrino wave functions and interference between lepton wave functions and nuclear wave functions bring different consequences for different decay branches. For decay branch I, on one hand the nuclear radius is small and the change of the electron wave functions are small inside the nucleus, on the other hand the Q value is small, the exact neutrino wave function can well be approximated by the long wave length approximation, so altogether, as for the point charge case, the differential width of the SA case and exact case are close, the SA and neutrino SA brings a correction less than 1\textperthousand. For decay branches II, as we have seen above, the decay happens near the surface, therefore SA works quite well, but due to it large Q value, the differential decay width at low electron energy has been largely reduced due to the adoption of exact neutrino wave function, this then distorts the $\beta$-spectra, brings accountable corrections as large as 1\% at the head of normalized $\beta$-spectra, this however has not been accounted before. For decay branches III and IV with large Z, exact accounting of the radial dependence of electron and neutrino wave functions gives rise to large enhancements to differential decay widths compared to SA, the correction is of the order of several percents. They also change the shape of the $\beta$-spectra, less strength will be shifted towards low electron energy region compared to SA. One finds that the corrected spectra lies always between the SA and FF curves.

Next I discuss the effects from different charge distributions. One finds that this effect is pretty small, for branches I and II, no visible difference can be found, the actual deviation is less than 1\textperthousand; for branches III and IV, the effects is larger but not enough to produce any real difference. As a consequence, for the normalized spectra, one then could barely distinguish among results from different charge distributions. However, one must be cautious to draw any conclusions before a thorough investigation on the effects of microscopic charge distributions, since what we use are obtained from electron nucleus scattering experiments and their forms are not far away from the uniform distribution, there may exist the probability that the actual nuclear charge distribution deviates large from these assumption in specific regions especially those with excess neutrons appearing possibly in reactors. 

\subsection{Estimations of errors}
\begin{figure*}
\includegraphics[scale=0.65]{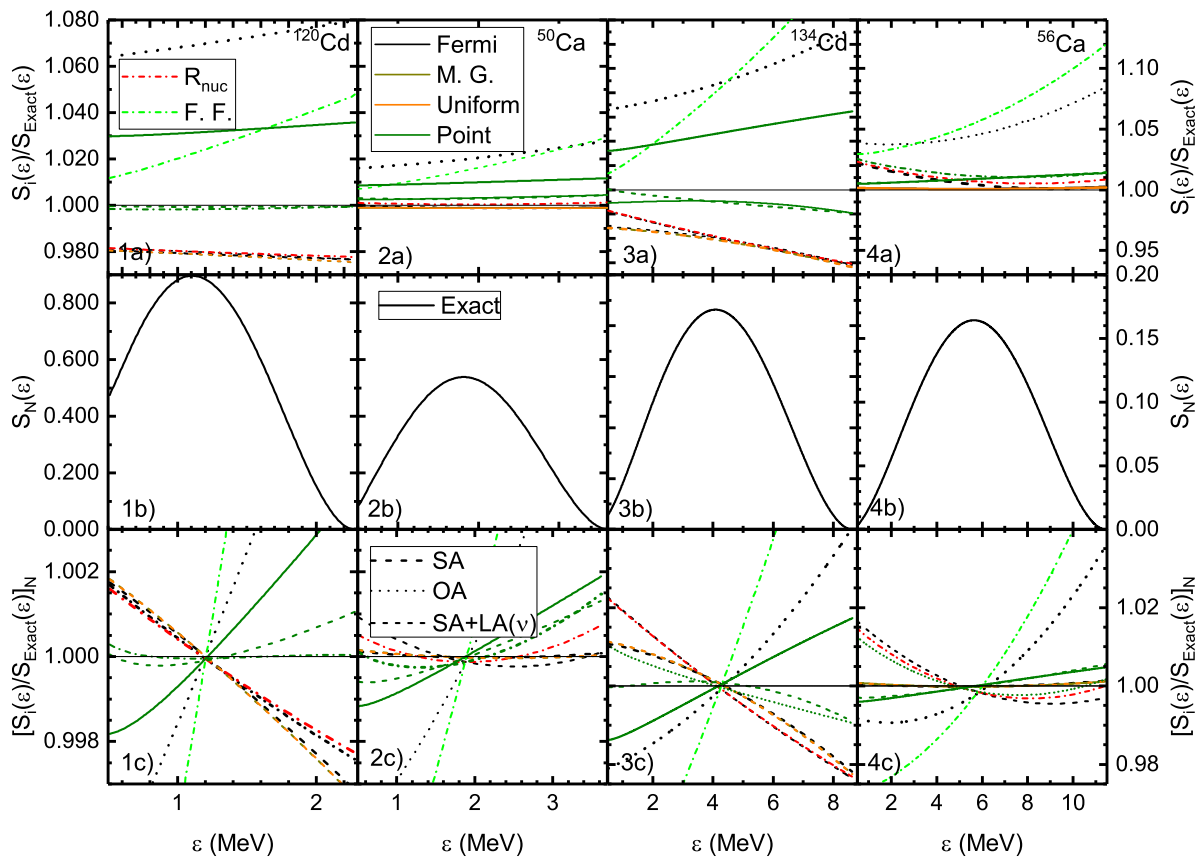}
\caption{(Color online) The errors from various approximations, for panel b's, the exact spectra for various channels are shown. The legends are as follows: in panel 1a) $R_{nuc}$ refers to SA using the empirical nuclear radius, F. F. refers to Fermi factor; in panel 2a) different charge distributions as indicated in text are shown; in panel 2b) 'Exact' refers to exact treatment of $\beta$-decay; in panel 2c), different line styles for different approximations are listed.}
\label{err}
\end{figure*}
In this section, I will make the semi-quantitative estimations of the errors from various approximations adopted in actual simulation. For the lepton part, accurate solutions with specific central potential can be achieved, but for nuclear structure part, we still have difficulties to make such conclusions. And an exact analysis of the errors are therefore difficult, but one could get general impression on how the nuclear structure will affect $\beta$-spectra. From the expressions above for the half-lives one finds that the they are affected by the electron wave functions and weak interactions between leptons and nucleons at the tree level order. In this work, I use the empirical electric charge distributions, therefore the effects of nuclear structure affects only the weak interactions and actually the weak charge distributions.

\subsubsection{Surface approximations {\it vs.} origin approximations}
 In analytical calculations, people usually use the results at the nuclear surface or at the origin. In this part, I will check which one is better and also which of the choice of nuclear radius helps to solve the problem. Taking a constant value of electron WF(either at the nuclear surface or center) could well separate the lepton from the nuclear structure, this is good for estimations of $\beta$- or neutrino- spectra. However, we need to understand how such approximation produce the errors to the actual spectra since in most cases, the measurements of branches single decay branches are difficult.
 
From fig.\ref{err}, we see that using wave functions at the origin brings a large deviation to the actual differential and total decay width. This again suggests that allowed decay is more likely to happen near the nuclear surface. SA in some occasions are close to actual decays especially for those decays dominated by single particle decays near surface. But in general, it will under predict the decay-widths for some branches here.

While SA is close to the results, origin approximation (OA) gives too large deviations, it gives a high over-prediction for decay width from several percent for low Q values to more than 10\% to large Q values. And we find that in some cases, its deviations are close to that of FF. And for the spectra, I find that OA behaves better than FF. Together with SA, I find that they set the boundary for the spectra, the actual decay branches will lie very likely in between and closer to SA curve.

I also want to mention here the choice of the surface, or namely the nuclear radius. One choice is the empirical one used commonly for estimating the various bulk properties of nuclei $R_{nuc}=1.2A^{1/3}$fm, and another is the so called Elton formula extract from electron-nucleus scattering experiments. These two radius are close to each other for nuclei in current work and Elton radii are known to be generally smaller than empirical one for A smaller than roughly 140 to 150. From fig.\ref{err} we see that such a difference in radii would cause a error for about several \textperthousand  for Ca isotopes and negligible for Cd isotopes. It can never become a major error for estimations of $\beta$-spectra, much less than the error caused by SA itself.

From above analysis, I find that using SA, one can get the lower boundary of differential decay width, and how large the deviation of actual decay width to this limit depends on the detailed nuclear structure of the parent and daughter nuclei. This is actually the uncontrolled errors from current nuclear structure calculations since an error estimation of there theories are difficult.  

\subsubsection{Neutrino wave function}
At the leading order, the neutrinos propagate in the space with the form of plane wave. And in usual calculations, the long-wave length approximation is used and the exact neutrino wave is not well considered. In above section for the point charge case, for small Q values, the errors are not significant. But for large Q values, we see that the spectra being distorted heavily.

In the case with SA of electron waves, the full consideration of neutrino wave functions will reduce the width at the low electron energy end and slightly enhance the high energy end, the reason of such behavior has been explained in the section of PCWF. And for the consideration of errors, I observe for branches II and IV the significant effect, however for branches I and III, this effect is not obvious due to their small Q values. 

Such behavior will affect the spectra only for branches with sufficient large Q values from above analysis. Distortions of the curves are observed for corresponding cases and naturally, they are heavily related to the values of Q. The errors become visible only for large Q values, on the other hand large Z will also increase the error as larger Z always corresponds to a larger nuclear radius $R$. For certain cases, the neutrino LA may become an important source of errors.

\subsubsection{Charge distributions}

Using the electron PCWF, we will get overestimated differential decay widths, this is due to the fact that such potential shifted the wave functions outwards compared to the finite volume charge potential. And with the increased electron energies, this deviation grows larger. With explicit integrations of PCWF, the intensity of $\beta$-spectrum is shifted upwards. If SA and $\nu$LA are imposed, the reduction of these approximations will compensate the enhancement of PCWF to the differential widths. This leads to cases where the PCWF and SA together produce the decay width close to the actual decay width. This suggests that such combinations can be used for estimation of better half-lives without the explicit solution of the Dirac equation with finite charge distribution size.

For electron wave functions from electric charges distributed in finite volume, the explicit integration give us negligible errors less than 1\textperthousand for the Uniform and Modified Gaussian distributions compared to the Fermi distribution we adopted as actual charge distribution, since the small magnitude of errors with these charge distribution, I omit them in the figure. In the future, more realistic electric charge distributions from nuclear microscopic calculations may be needed to clarify whether the charge distributions could produce significant errors.

From above analysis, I find that when simulating the reactor $\beta$-spectra using the summation method, if no exact nuclear structure information is presented, there will be invertible errors with magnitude depending on Z and Q  for most decay branches. With present situations of nuclear structure calculations, all the simulated spectra will suffer from this error and they can not be get rid of in the foreseeable future. While the choices of finite volume charge distributions seem to be not so important for spectra simulation, the simplest uniform distribution could give us enough accuracy.

The constraints of the errors can alway be done. The spectra with SA can be served for this purpose, this can be seen from the non-solid curves in fig.\ref{err}.  Where we see that F.F. or Spectra with OA give the upper limit for low electron energy and lower limit for high electron energies and SA with FSWF gives the opposite boundaries.

\section{Conclusion and outlook}
In this work, with modern electron wave functions and many-body approaches, I examined how the errors will different approximations bring. The results shows that the analytical calculations are generally good for calculations of decay width and spectra with certain accuracy, but if one needs accuracy beyond $1\%$, then we need to bring in the exact electron wave functions. And the lack of high-precision nuclear structure theory will bring the inevitable errors to the simulation for decay branches with large Q-values. Also, under such a frame, the effect of high order terms in electromagnetic and weak decay theory needs further investigation as well as the screening effect of the atomic orbital electrons. The corresponding modification to neutrino spectra from this work can be easily accessed by reversing the conclusions for $\beta$-spectra. Meanwhile I find that the errors from all these approximations are way smaller than that brought by replacing first forbidden decay by allowed decay\cite{FB15}.
 
 \section*{Acknowledgement}
This work is supported in part by the National Key Research and Development program (MOST 2016YFA0400501) from the Ministry of Science and Technology of China and Nation science foundation of China under grant No. 11505078 and 1164730. I would like to thank Prof. B. A. Brown for helps on shell model calculations. I would like also to acknowledge the useful discussions and helpful suggestions from Prof. L. G. Jiao and Dr. Y. F. Li. I would thank Mr. H. Y. Ma and Q. L. Lu for double check of the accuracy of {\it Radial }package.

%\bibliographystyle{apsrev}
%\bibliography{bspct}

\end{document}